\documentclass[twoside]{hs04}
\def\runauthor{O.V.Utyuzh, G.Wilk, Z.W\l odarczyk}
\def\shorttitle{Multiparticle production processes...}
\begin{document}

\title{MULTIPARTICLE PRODUCTION PROCESSES\\
FROM THE NONEXTENSIVE POINT OF VIEW
}

\author{O.V.UTYUZH and G.WILK}
\address{The Andrzej So\l tan Institute for Nuclear Studies, Zd-PVIII\\
Ho\.za 69, 00-681 Warszawa, Poland\\
E-mails: utyuzh@fuw.edu.pl and wilk@fuw.edu.pl}
\author{Z.W\L ODARCZYK}
\address{Institute of Physics, \'Swi\c{e}tokrzyska Academy\\
\'Swi\c{e}tokrzyska 15, 25-406 Kielce, Poland\\
E-mail: wlod@pu.kielce.pl}
\maketitle

\abstracts{We look at multiparticle production processes from the
nonextensive point of view. Nonextensivity means here the systematic
deviations in exponential formulas provided by the usual statistical
approach for description of some observables like transverse momenta
or rapidity distributions. We show that they can be accounted for my
means of single parameter $q$ with $|q-1|$ being the measure of
nonextensivity. The whole discussion will be based on the information
theoretical approach to multiparticle processes proposed by us some
time ago.
}

\section{Introduction}

The multiparticle production processes are usually approached first
by means of statistical models \cite{Stat} in order to make quick
estimations of such parameters as "temperature" $T$ or "chemical
potential" $\mu$. The quotation marks reflect the controversy in what
concerns the applicability of such terms in the realm of hadronic
production. Actually, as has been pointed long time ago \cite{LVH},
the fact that out of $N$ measured particles we usually observe only
one of them is enough to make the relevant distribution (in
transverse momentum $p_T$ or in rapidity $y$) being exponential:
\begin{equation}
f(X) \sim \exp\left[ - \frac{X}{\Lambda} \right] . \label{eq:EXP}
\end{equation}
This is because the remain $N-1$ particles act as a "heath bath" to
the observed one and their action can be therefore summarized by a
single parameter $\Lambda$. Identification of $\Lambda$ with
temperature $T$ means that we assume that our system is in thermal
equilibrium. However, in reality such "heath bath" is not infinite
and not homogenous (i.e., it can contain domains which differ between
themselves). In such case single parameter $\Lambda$ is not
sufficient to summarily describe its influence on the particle under
consideration. One can regard it as being $X$-dependent and use the
first two terms of its expansion around some value $X_0$ (instead of
only $X_0$): $\Lambda \simeq \Lambda_0 + a\cdot (X-X_0)$. In this
case distribution (\ref{eq:EXP}) becomes
\begin{equation}
f_q(X) \sim \exp_q\left[ - \frac{X}{\Lambda_0} \right] \stackrel{def}{=}
     \left[1 - (1-q)\frac{X-X_0}{\Lambda_0}\right]^{\frac{1}{1-q}}
     \qquad {\rm where}\qquad q=1 + a .
        \label{eq:EXPq}
\end{equation}
In case when $X$ is energy $E$ and $\Lambda$ is temperature $T$ the
coefficient $a$ is just the reverse heat capacity, $a=1/C_V$
\cite{Alm}. Such phenomenon is called {\it nonextensivity} and
with $q$ being the {\it nonextensivity parameter}. Notice that
for $q\rightarrow 1$ (or for $a\rightarrow 0$) $f_q(X)$ becomes
$f(X)$.

In next Section we shall look at this problem from the information
theory of view using Shannon and Tsallis information entropies.
Confrontation with experimental data will be provided in Section 3.
Section 4 will contain summary and conclusions.

\section{Information theory and multiparticle production processes}

Let us consider situation, which happens quite often in physics
realm. Suppose that experimentalist provided us with some new data.
Immediately these data are subject of interest to a number of
theoretical groups, each of them quickly proposing their own
distinctive and unique (in what concerns assumptions and physical
concepts) explanation. Albeit disagreeing on physical concepts they
are all fairly successful in what concerns fitting data at hand. The
natural question which arises is: {\it which of the proposed models
is the right one?} The answer is: {\it to some extend - all of them}!
This is because experimental data are providing only limited amount
of information and all models are simply able to reproduce it. To
quantify this reasoning one has to define the notion of information
and to do this, one has to resort to Information Theory
(IT)\footnote{The complete list of references concerning IT relevant
to our discussion here and also providing some necessary background
can be found in \cite{MaxEnt,MaxEntq}.}. It is based on Shannon
information entropy
\begin{equation}
S = - \sum_i \, p_i\, \ln p_i \, , \label{eq:Shannon}
\end{equation}
where $p_i$ denotes probability distribution of interest. Notice that
the least possible information, corresponding to the equal
probability distribution of $N$ states (i.e., $p_i = 1/N$), results
in maximal entropy, $S=\ln N$. The opposite situation, when only one
state is relevant, i.e., $p_k=1$ and $p_{i\neq k}=0$, results in the
minimal entropy, $S=0$. Usually one always has some {\it a priori}
information on experiment, like conservation laws and results of
measurements, which is represented by a set of known numbers,
$\langle R_k\rangle$. One is then seeking probability distribution
$\{p_i\}$ such that:
\begin{itemize}
\item it tells us {\it the truth, the whole truth} about our
experiment, what means that, in addition to be normalized, $\sum_i p_i
=1$, it reproduces the known results, i.e.,
\begin{equation}
\sum_i\, p_i\, R_k(x_i)\, =\, \langle R_k\rangle\, ;
\label{eq:constraints}
\end{equation}
\item it tells us {\it nothing but the truth} about our experiment,
i.e., it conveys {\it the least information} (only the information
contain in this experiment).
\end{itemize}
To find such $\{p_i\}$ one has to {\it maximize} Shannon entropy under
the above constraints (therefore this approach is also known as
MaxEnt method). The resultant distribution has familiar exponential
shape
\begin{equation}
p_i\, =\, \frac{1}{Z}\, \cdot \exp\left[ - \sum_k\lambda_k\cdot
R_k(x_i) \right]\, . \label{eq:MEp}
\end{equation}
Although it looks identical to the "thermal-like" (Boltzmann-Gibbs)
formula (\ref{eq:EXP}) there are {\it no free parameters} here
because $Z$ is just normalization constant assuring that $\sum p_i
=1$ and $\lambda_k$ are the corresponding lagrange multipliers to be
calculated from the constraint equations
(\ref{eq:constraints})\footnote{Notice that using the entropic
measure $S\, =\, \sum_i\left[p_i \ln p_i\, \mp\, \left(1\, \pm\,
p_i\right) \ln \left(1\, \pm\, p_i \right) \right]$ (which, however,
has nothing to do with IT) would result instead in Bose-Einstein and
Fermi-Dirac formulas: $p_i\, =\, (1/Z)\cdot
\left[\exp[\beta(\varepsilon_i - \mu)] \mp 1\right]^{-1}$,
where $\beta$ and $\mu$ are obtained solving two constraint equations
given, respectively, by energy and number of particles conservation
\cite{TMT}. It must be also stressed that the final functional form
of $p_i$ depends also on the functional form of the constraint functions
$R_k (x_i)$. For example, $R(x) \propto \ln (x)$ and $\ln (1-x)$
type constrains lead to $p_i \propto x_i^{\alpha}(1-x_i)^\beta$
distributions.}.

It is worth to mention at this point \cite{MaxEnt} that the most
probable multiplicity distribution $P(n)$ in the case when we know
only the mean multiplicity $\langle n\rangle$ and that all particles
are distinguishable is geometrical distribution $P(n) = \langle
n\rangle^n/(1+\langle n\rangle)^{(n+1)}$ (which is broad in the sense
that its dispersion is $D(n) \sim \langle n\rangle$). Additional
knowledge that all these particles are indistinguishable converts the
above $P(n)$ into Poissonian form, $P(n) = \exp ( -\langle n\rangle
)/n!$, which is the narrow one in the sense that now its dispersion is
$D(n) \sim \sqrt{\langle n\rangle}$. In between is situation in which
we know that particles are grouped in $k$ equally strongly emitting
sources, in which case one gets Negative Binomial distribution
\cite{NBin} \footnote{It is straightforward to check that Shannon
entropy decreases from the most broad geometrical distribution
towards the most narrow Poissonian distribution.} $P(n) =
\Gamma(k+n)/\left[\Gamma(n+1)\Gamma(k)\right]\cdot \left(
\frac{k}{\langle n\rangle }\right)^k/\left[ 1 + \frac{k}{\langle
n\rangle}\right]^{k+n}$.

The other noticeably example concerns the use of IT to find the
minimal set of assumptions needed to explain all multiparticle
production data of that time \cite{Chao}. It turned out that all
competing models like: multi-Regge, uncorrelated jet,
thermodynamical, hydrodynamical etc., shared (in {\it explicit} or
{\it implicit} manner), two basic assumptions:
\begin{itemize}
\item[$(i)$] that only part $W=K\cdot\sqrt{s}$ ($0<K<1$) of the
initially allowed energy $\sqrt{s}$ is used for production of
observed secondaries (located in the center part of the phase space);
in this way {\it inelasticity} $K$ found its justification \cite{Kq},
it turns out that $K\sim 0.5$;
\item[$(ii)$] that transverse momenta of produced particles are
limited and the resulting phase space is effectively one-dimensional
(dominance of the longitudinal phase space).
\end{itemize}
All other assumptions specific for a given model turned out to be
just spurious\footnote{The most drastic situation was with the
multi-Regge model in which, in addition to the basic model
assumptions, two purely phenomenological ingredients have been
introduced in order to get agreement with experiment:  $(i)$ energy
$s$ was used in the scaled $(s/s_0)$ form (with $s_0$ being free
parameter, this works the same way as inelasticity) and $(ii)$ the so
called "residual function" factor $e^{\beta\cdot t}$ was postulated
($t=-(p_i-p_j)^2$ and $\beta$ being a free parameter) to cut the
transverse part of the phase space. Therefore $s_0$ and $\beta$ were
{\it the only important} parameters whereas all other model
parameters were simply irrelevant.}.

Suppose now  that some new data occur which disagree with the
previously established form of $\{p_i\}$. In IT approach it simply
signals additional information to be accounted for. This can
be done either by adding a new constraint (resulting in new
$\lambda_{k+1}$) or by checking whether one should not use
$\exp_q(...)$ rather than $\exp(...)$. This brings to IT the notion
of nonextensivity introduced before and is connected with freedom in
choosing the form of information entropy. The point is that there are
systems which experience long range correlations, memory effects,
which phase space has fractal character or which exhibit some
intrinsic dynamical fluctuations of the otherwise intensive
parameters (making them extensive ones, like $T$ in the example
above). Such systems are truly {\it small} because the range of
changes is of the order of their size. In this case the Shannon
entropy (\ref{eq:Shannon}) is no longer good measure of information
and should be replaced by some other measure. Out of infinitely many
possibilities \cite{CT,ENT} we shall choose Tsallis entropy defined
as
\begin{equation}
S_q\, =\, - \frac{1}{1-q}\sum_i\, \left( 1 - p_i^q \right),
\label{eq:Tsallis}
\end{equation}
which, as can be easily shown, goes over to Shannon form
(\ref{eq:Shannon}) for $q\rightarrow 1$. To get the $\{p_i\}$
one proceeds in the same way as before but modifying the constraint
equation in the following way,
\begin{equation}
\sum_i\, \left[p_i\right]^q\, R_k(x_i)\, =\, \langle R^{(q)}_k\rangle
. \label{eq:q_constr}
\end{equation}
This leads to formally the same formula for $p_i=p^{(q)}_i$ as in eq.
(\ref{eq:MEp}) but with $Z \rightarrow Z_q$ and $\exp (...)
\rightarrow \exp_q (...)_q$. (For our purpose this method is
sufficient and there is no need to use the more sophisticated
approach exploring the so called escort probabilities formalism, see
\cite{Kq}). Because such entropy is nonextensive, i.e., $S_{q(A+B)} =
S_{qA}+S_{qB} + (1-q)S_{qA}\cdot S_{qB}$ (see \cite{CT} for more
details), the whole approach takes name nonextensive (Tsallis)
statistics. Nonextensivity parameter $q$ accounts summarily for the
all possible sources of nonextensivity mentioned before, however,
here we shall be interested here only in intrinsic fluctuations
present in the system. This is because in this case, as was
demonstrated in \cite{WW}, parameter $q$ is measure of such
fluctuations, namely for system described by eq. (\ref{eq:EXPq}) one
has that \footnote{Strictly speaking in \cite{WW} it was shown only
for fluctuations of $1/\Lambda$ given by gamma distribution. However,
it was soon after generalized to other form of fluctuations and the
word {\it superstatistics} has bee coined to describe this new
phenomenon \cite{CB}.}
\begin{equation}
q\, =\, 1\, \pm\,
         \left[
             \left\langle \left(1/\Lambda\right)^2\right\rangle\,
             -\,
             \left\langle 1/\Lambda \right\rangle^2
        \right] /
             \left\langle 1/\Lambda \right\rangle^2 . \label{eq:Q}
\end{equation}
In what follows we shall argue that multiparticle production data on
both transverse momentum $p_T$ and rapidity $y$ distributions can be
interpreted as showing effects of such fluctuations.

\section{Confrontation with experimental data}

One should stress that what we present here is not a model but only
{\it least biased} and {\it most probable} (from the point of view of
information theory) distributions describing available data and
accounting for constraints emerging either from the conservation laws
or from some dynamical input (which in this way is subjected to
confrontation with data). Suppose that we have mass $M$, which
hadronizes into $N$ secondaries and hadronization is taking place in
$1-$dimensional longitudinal phase space. Dynamics is hidden here in
the fact that decay is into $N$ particles and that transverse momenta
of particles are limited so one can use their transverse mass $\mu_T$
(suppose that we are not interested for a moment in details of the
transverse momentum distribution). Once this is settled IT gives us
clear prescription what to do:
\begin{itemize}
\item Look for single particle distribution in longitudinal phase
space taken as $y-$space and written as suitable probability
distribution,
\begin{equation}
f_N(y)\, =\, \frac{1}{N}\cdot \frac{dN}{dy}\, =\, \frac{1}{\sigma_N
N}\int d^2p_T \frac{Ed\sigma}{d^3p} . \label{eq:defy}
\end{equation}
\item Maximize the following entropy functional
\begin{equation}
S\, =\, - \int^{Y_M}_{-Y_M}dy\, f_N(y)\cdot \ln\left[f_N(y)\right],
\label{eq:enty}
\end{equation}
under conditions that $f_N(y)$ is properly normalized and that one
conserves energy (because of the symmetry of our problem the momentum
will be satisfied automatically; no more information will be used):
\begin{equation}
\int^{Y_M}_{-Y_M}\!\!\!f_N(y) =1\quad {\rm and} \quad
\int^{Y_M}_{-Y_M}\!\!\!dyE\cdot f_N(y) = \mu_t \int^{Y_M}_{-Y_M}\!\!\!
dy f_N(y)\cosh y = \frac{M}{N}. \label{eq:constry}
\end{equation}
\end{itemize}
As result one obtains that the most probable (and least biased)
rapidity distribution is
\begin{equation}
f_N(y)\, =\, \frac{1}{Z}\cdot \exp\left[ - \beta\cdot \cosh y
\right]\quad {\rm where}\quad
Z = \int^{Y_M}_{-Y_M}\!\!\!dy \exp\left[ - \beta\cdot \cosh y \right]
\end{equation}
and where $\beta=\beta (M,N)$ is obtained by solving eq.
(\ref{eq:constry}). The $y-$space is limited to
$$ y\in(-Y_M, Y_M),\quad Y_M = \ln
\left\{\frac{M'\mu_t}{2\mu_T}\left[1+\left( 1-
\frac{4\mu_T^2}{M'^2}\right)^{1/2}\right]\right\},~~
M'=M-(N-2)\mu_T.$$
The following points should be remembered:
\begin{itemize}
\item For $N=2$ both particles must be at the ends of the phase
space, i.e., $f_2(y) = \frac{1}{2}\left[\delta(y-Y_M) +
\delta(y+Y_M)\right]$, therefore formally $\beta (M,2) = -\infty$.
\item Between $N=2$ and $N\simeq N_0=2\ln N_{max}$ (where
$N_{max}=M/\mu_T$) $\beta$ is negative and approaches zero {\it only}
for $N = N_0$. Therefore only for $N=N_0$ we have $f_N(y)=const$,
i.e., the famous Feynman scaling\footnote{It was jus a coincidence
that at ISR energies this condition was satisfied. But such behaviour
of $N$ as function of energy was only transient phenomenon and there
will never be scaling of this type, notwithstanding all opinions to
the contrary heard from time to time.}.
\item For $N>N_0$ additional particles have to be located near the
middle of phase space to in order to minimize energy cost of their
production. As result $\beta > 0$ now, in fact (see \cite{MaxEnt} for
details) for some ranges of $<E>=M/N$ quantity $\bar{\beta} = \beta
<E>$ remains approximately constant. Needles to say that for
$N\rightarrow N_{max}$ {\it all particles} have to stay as much as
possible at the center and therefore $\beta(M,N\rightarrow N_{max})
\rightarrow +\infty$ whereas $f_{N\rightarrow N_{max}}(y) \rightarrow
\delta(y)$.
\end{itemize}
\vspace{-10mm}
\begin{figure}[ht]
  \begin{minipage}[ht]{50mm}
    \centerline{
        \epsfig{file=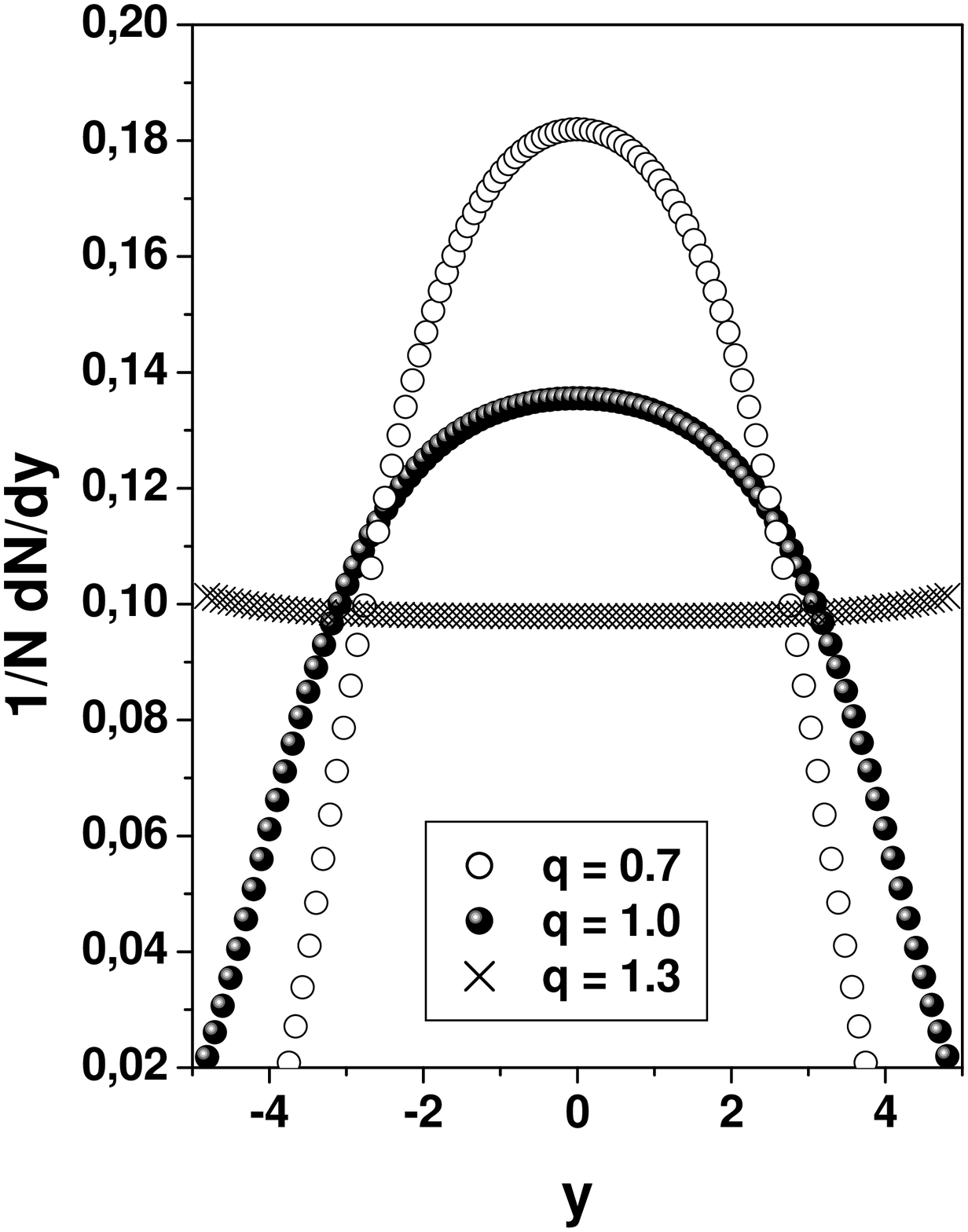, width=40mm}
     }
  \end{minipage}
\hfill
  \begin{minipage}[ht]{70mm}
    \centerline{
        \epsfig{file=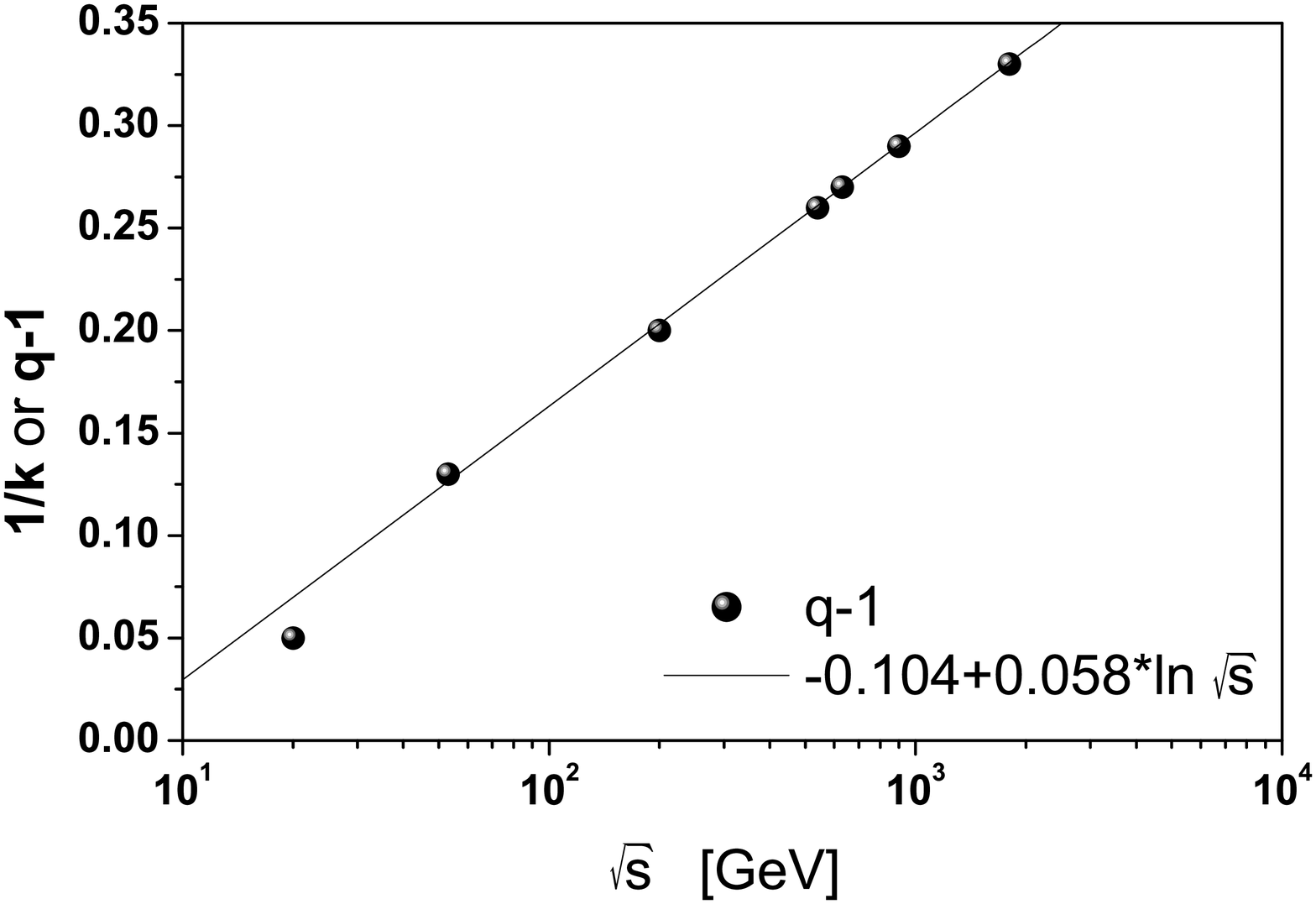, width=70mm}
     }
  \end{minipage}
\vspace{-8mm}
  \caption{
\footnotesize {Left panel: examples of rapidity distributions for
different nonextensivities. Right panel: the energy dependence of the
NB parameter $1/k$ (line) together with results for the
nonextensivity $|q-1|$ obtained when fitting $pp$ and $\bar{p}p$ data
at different energies (points) - see the top-left panel of Fig.
\ref{fig:Fig2}.}}
 \label{fig:Fig1}
\end{figure}
\vspace{-5mm}

For nonextensive approach the only changes to be performed is to
replace $\exp(...)$ by $\exp_q(...)_q$ in the way explained before.
In addition $N_0 \rightarrow N^{(q)}_0 = \left( 2 \ln
N_{max}\right)^q$. The changes in the shapes of $f_q(y)$ function for
different values of $q$ are shown in Fig. \ref{fig:Fig1}\footnote{In
general situation is bit more involved, see ,for example, \cite{Beck}
for more details. This, however, is out of scope of our
presentation.}. Notice that for $q>1$ one enhances tails of
distribution and at the same time reduced its height. For $q1$ the
effect is opposite (in this case there is kinematical constraint on
the allowed phase space, namely it must conform to the condition that
$1-(1-q)\beta_q\mu_T\cosh y \ge 0$. In fact in \cite{NuCim} we have
found that we can fit reasonably well data on $\bar{p}p$
\cite{ppData} with $q$ being the only parameter. Resultant $q<1$,
which was cutting off the available phase space was there playing
effectively the role of inelasticity of reaction mentioned
before\footnote{The aim of \cite{NuCim} was to provide the cosmic ray
physicist community a kind of justification of the empirical formula
they used, namely that $f(x) \propto (1-a\cdot x)^n$, where $x$ is
Feynman variable, $x=2E/\sqrt{s}$, and where $a$ and $n$ are free
parameters. They turned to be both given by $q$ only \cite{NuCim}.}.
\begin{figure}[ht]
  \begin{minipage}[ht]{57mm}
    \centerline{
        \epsfig{file=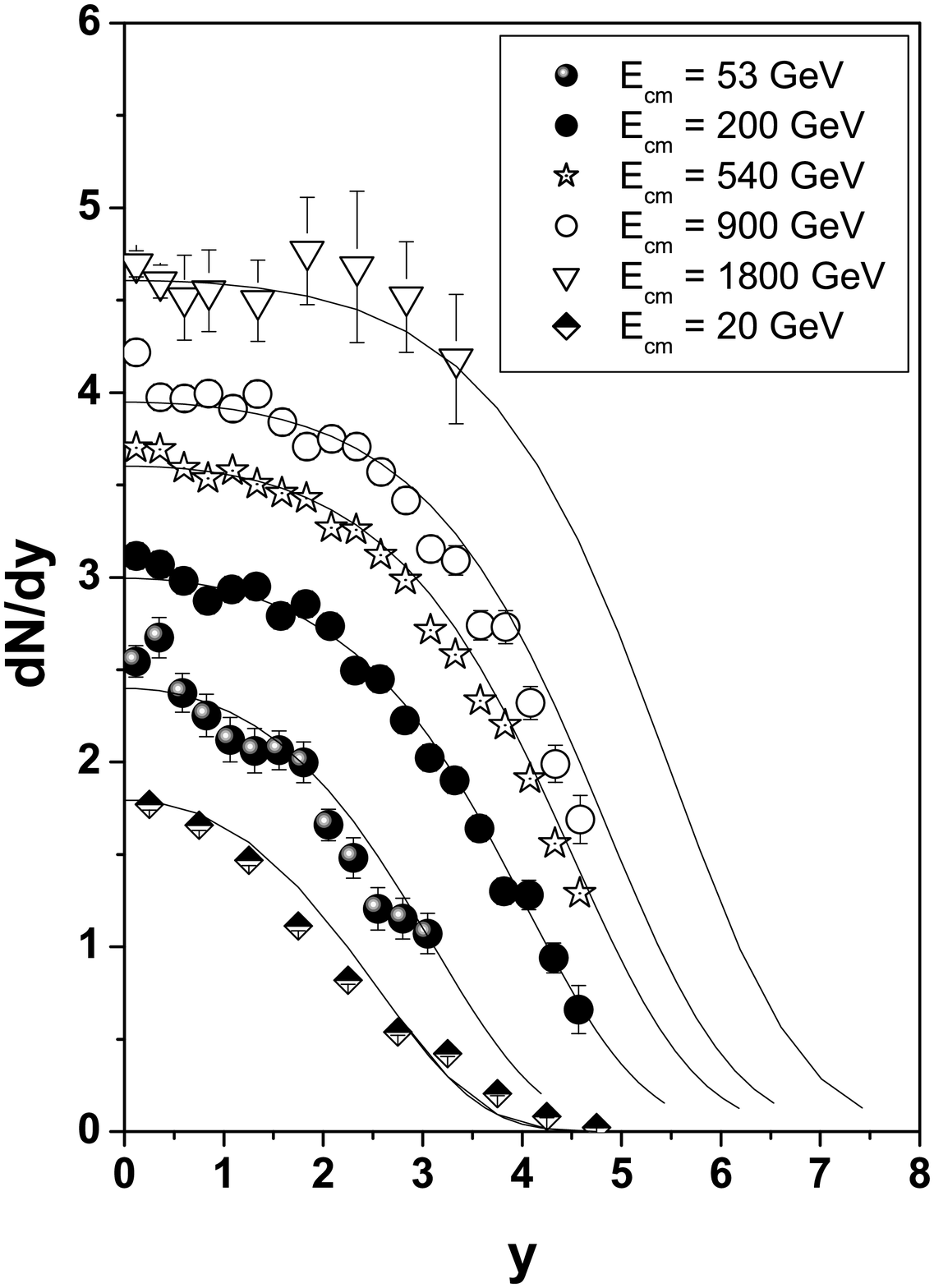, width=55mm}
     }
  \end{minipage}
\hfill
  \begin{minipage}[ht]{57mm}
    \centerline{
        \epsfig{file=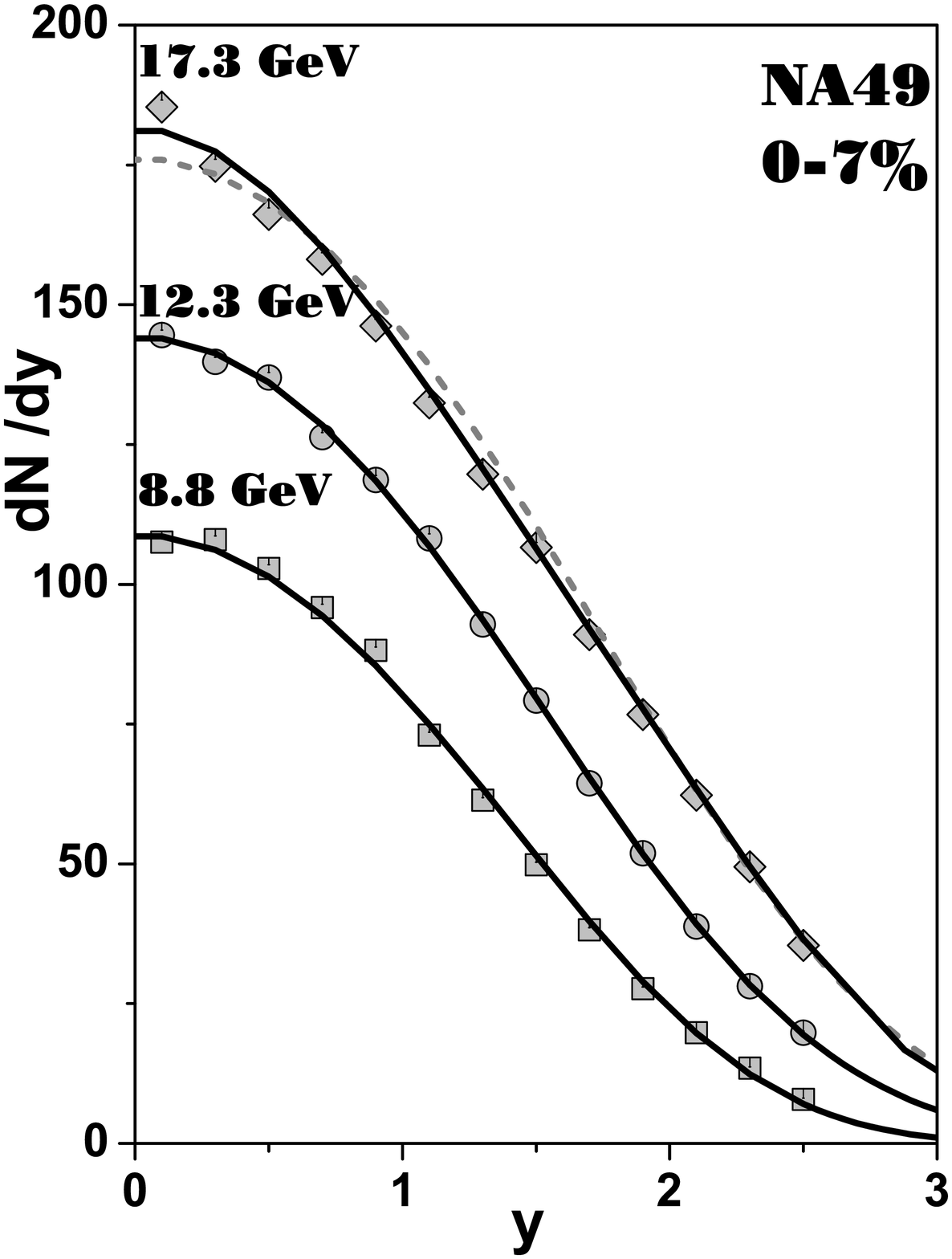, width=55mm}
     }
  \end{minipage}
\hfill
  \begin{minipage}[ht]{57mm}
    \centerline{
        \epsfig{file=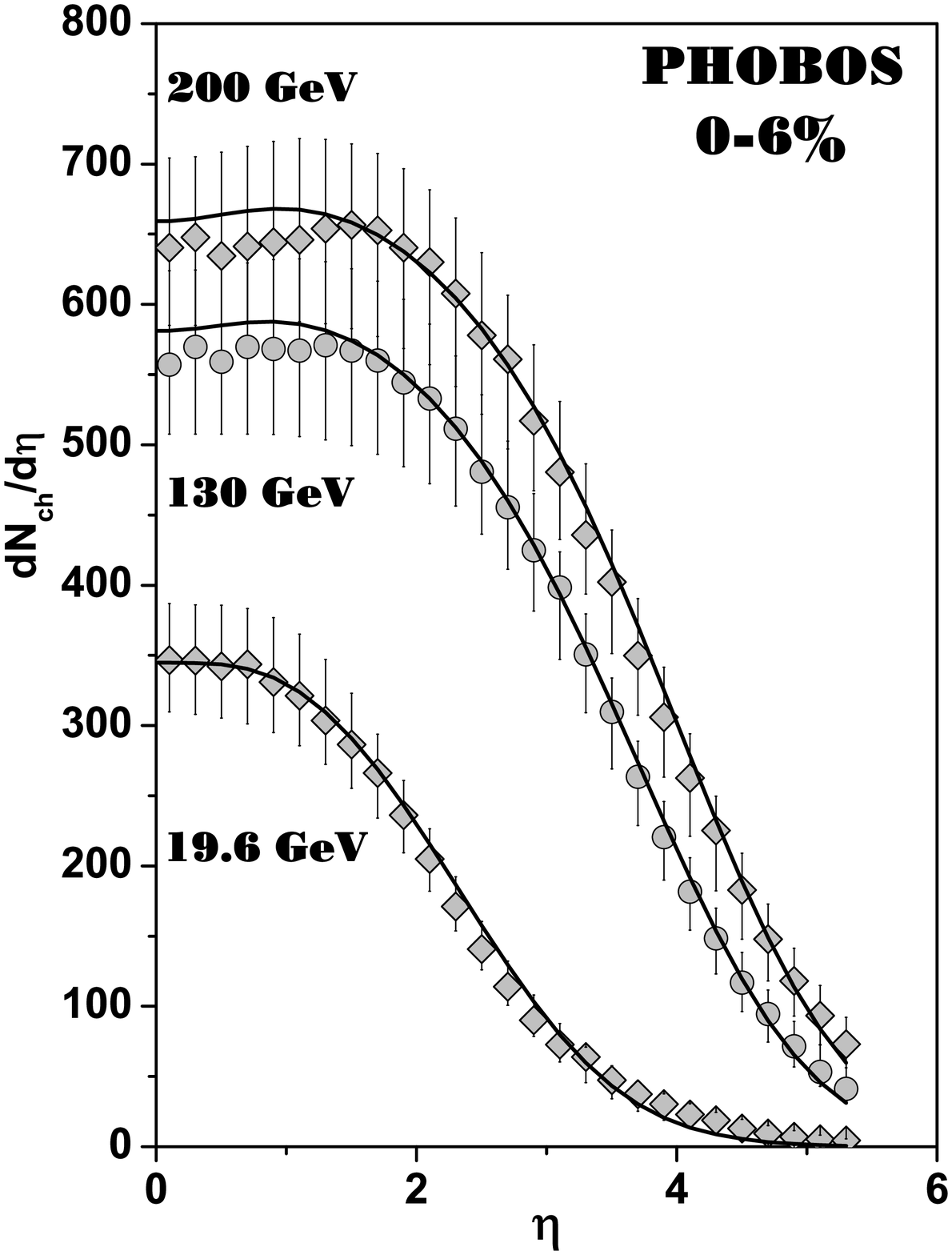, width=55mm}
     }
  \end{minipage}
\hfill
  \begin{minipage}[ht]{57mm}
    \centerline{
       \epsfig{file=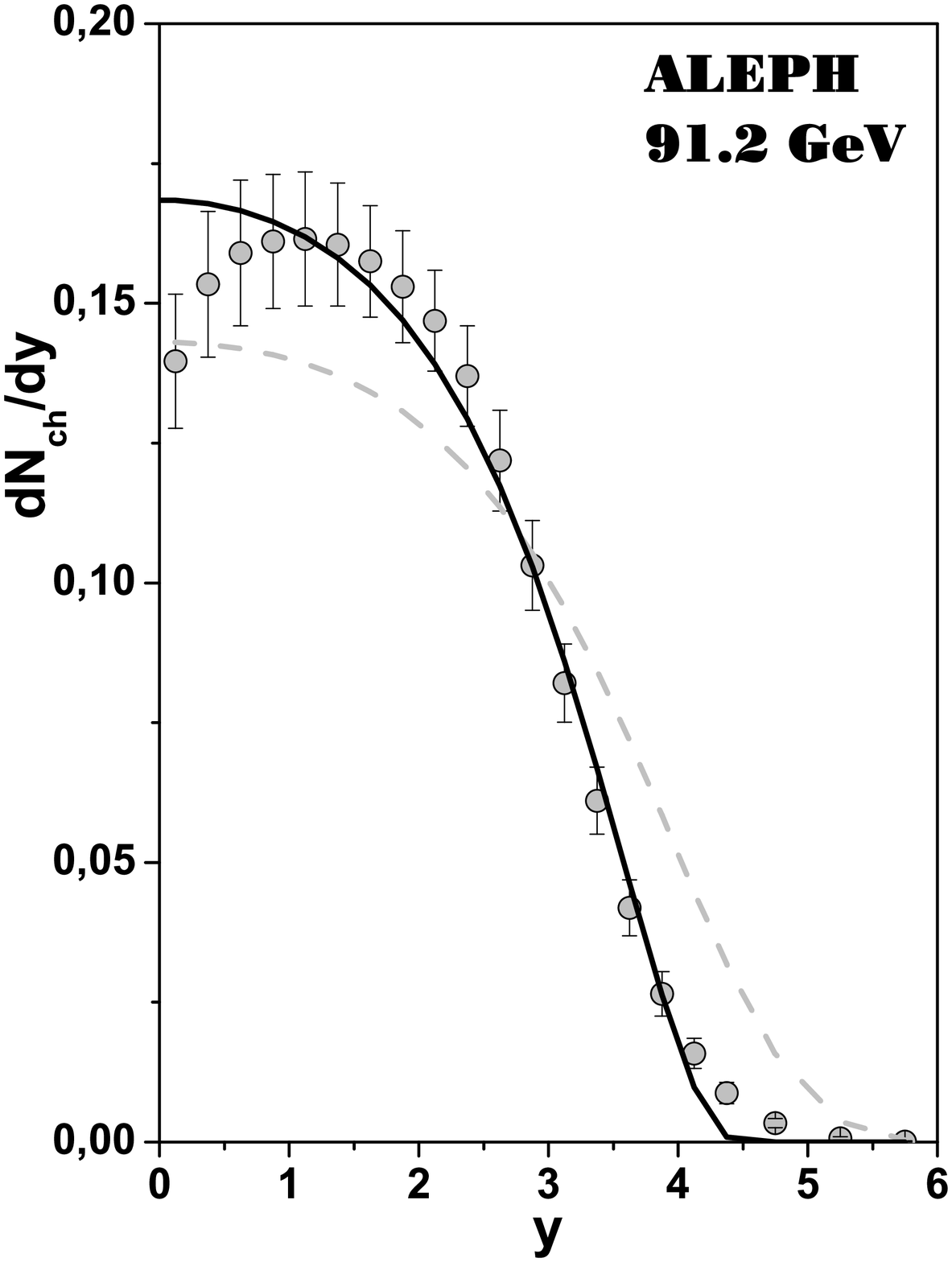, width=55mm}
     }
  \end{minipage}
  \caption{
\footnotesize {The most characteristic examples of comparison with
data on rapidity distributions for different reactions. Top-left
panel: the best fits to $pp$ and $\bar{p}p$ data
\protect\cite{ppData} using nonextensive approach (cf.
\protect\cite{Kq}). The resulting values of the $q$ parameters are
shown above in Fig. \ref{fig:Fig1}. Top-right panel: fits to NA49
data for negatively charged pions produced in central $PbPb$
collisions at different energies \protect\cite{DataAA} (cf.
\protect\cite{MaxEntq}). The best fit for $17.3$ GeV is, however, for
two extensive sources located at $y =\pm \Delta y = 0.83$ in rapidity
(total inelasticity in this case is $K=0.58$). Bottom-left panel:
fits to PHOBOS data for the most central $Au+Au$ collisions
\protect\cite{Data4} (cf., \protect\cite{Trends}). Bottom-right
panel: rapidity spectra measured in $e^+e^-$ annihilations at $91.2$
GeV \protect\cite{Data5} (dotted line is for $K_q=1$ and $q=1$
whereas full line is our fit with $K_q=1$ and $q=0.6$, see
\protect\cite{Trends,Nukleon}).}}
  \label{fig:Fig2}
\end{figure}

We repeated attempt to fit $\bar{p}p$ (and this time also $pp$) data
\cite{ppData} again but this time accountig properly for inelasticity
(which was now free parameter to be deduced from data) and also for
other possible fluctuations which hadronizing process offers us and
which were represented by parameter $q$. We shall not discuss here the
inelasticity issue (see \cite{Kq} for details and further references)
but concentrate on the $q$ parameter. In Fig. \ref{fig:Fig1} we show
the values of this parameter obtained for different energies compared
with the values of the parameter $k$ of Negative Binomial
distribution (NB) taken for different energies from existing data
\cite{NB}. The agreement is perfect what prompted us to argue that
parameter $q$ is accounting for a new bit of information, so far
unaccounted for, namely that in addition to the mean multiplicity
which was our input parameter there are other data \cite{NB} showing
the whole multiplicity distributions, in particular showing their NB
character. Actually this conjecture is supported by the observation
\cite{NBin} that NB can be obtained from Poisson distribution once
one allows for fluctuations in its mean value $\bar{n}$ proceeding
according to gamma distribution, namely
\begin{equation}
P(n)\, =\, \int_0^{\infty} d\bar{n} \frac{e^{-\bar{n}}\bar{n}^n}{n!}\cdot
         \frac{\gamma^k \bar{n}^{k-1} e^{-\gamma \bar{n}}}{\Gamma (k)} =
   \frac{\Gamma(k+n)}{\Gamma (1+n) \Gamma (k)}\cdot
   \frac{\gamma^k}{(\gamma +1)^{k+n}} \label{eq:PNBD}
\end{equation}
where $\gamma = \frac{k}{\langle \bar{n}\rangle}$ and $
\frac{1}{k} = \frac{\sigma^2(n_{ch})}{\langle n_{ch}\rangle^2}-
\frac{1}{\langle n_{ch}\rangle}$.
Assuming now that these fluctuations contribute to nonextensivity
defined by the parameter $q$, i.e., that $D(\bar{n}) = q-1$ one gets
that
\begin{equation}
q\, =\, 1\, +\, \frac{1}{k}, \label{eq:qk}
\end{equation}
what we do observe (cf. also \cite{Cern}).

The next two panels show comparison with $AA$ data taking at SPS
(NA49) and RHIC (PHOBOS) energies. The NA49 data can be fitted with
$q=1.2$, $1.16$ and $1.04$ going from top to bottom and so far we do
not have plausible explanation for these number. However, it should
be stressed that at highest energy two component extensive source is
preferable. The PHOBOS data with $q=1.27$, $1.26$ and $1.29$ going
from top to bottom. Extensive fits do not work here at all. The
example of $e^+e^-$ data clearly show that in this case there is some
new information we did not accounted for. In this case the whole
energy must be used but only $q<1$ fit (cutting-off a part of phase
space) can be regarded as a fair one. But even then we cannot obtain
minimum at $y=0$. It looks like we have two sources here separated in
rapidity (two jets of QCD analysis) but of no statistical origin
(rather connected with cascading) \cite{Nukleon}.

\begin{figure}[ht]
  \begin{minipage}[ht]{57mm}
    \centerline{
        \epsfig{file=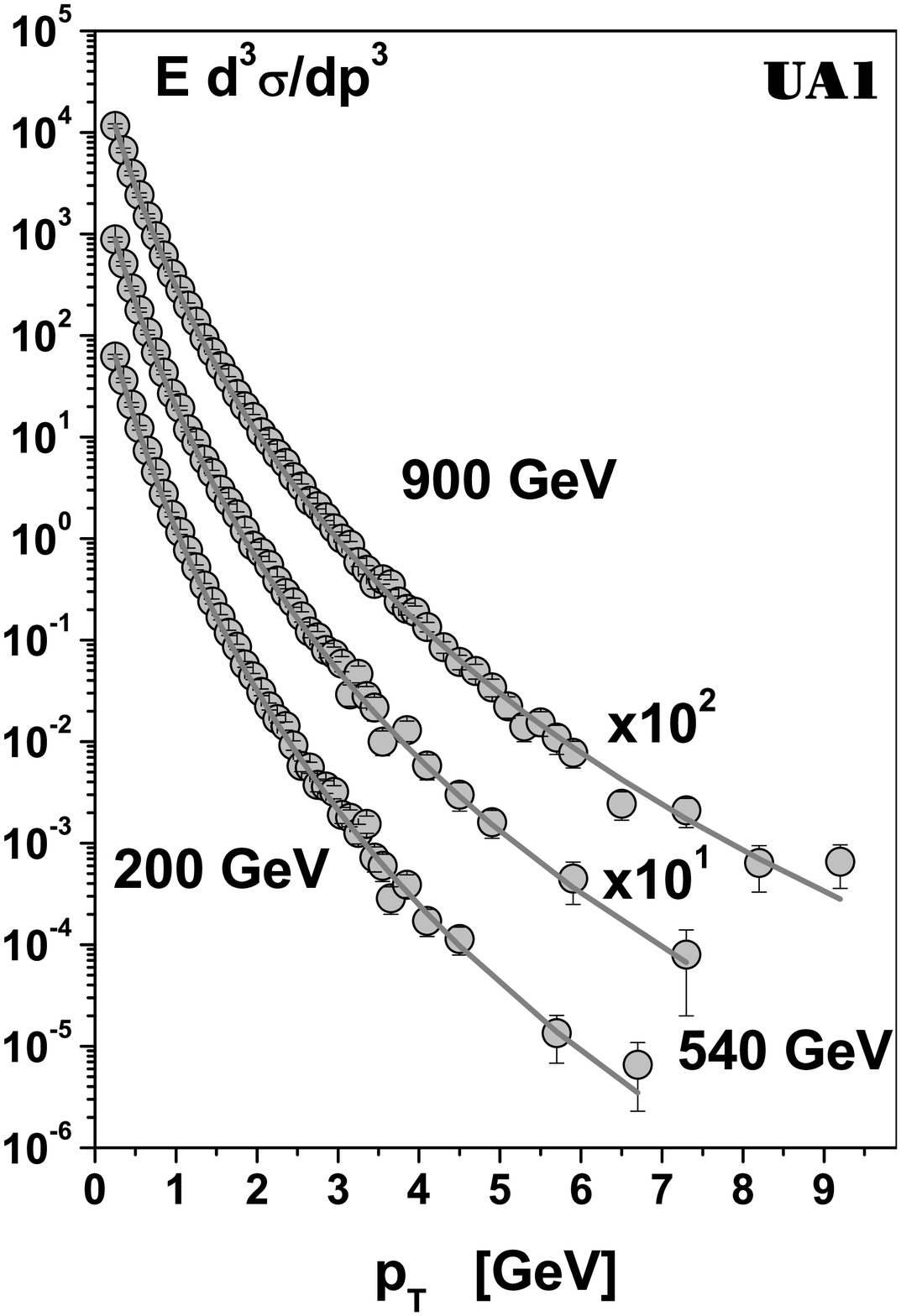, width=50mm}
               }
  \end{minipage}
\hspace{5mm}
  \begin{minipage}[ht]{57mm}
    \centerline{
        \epsfig{file=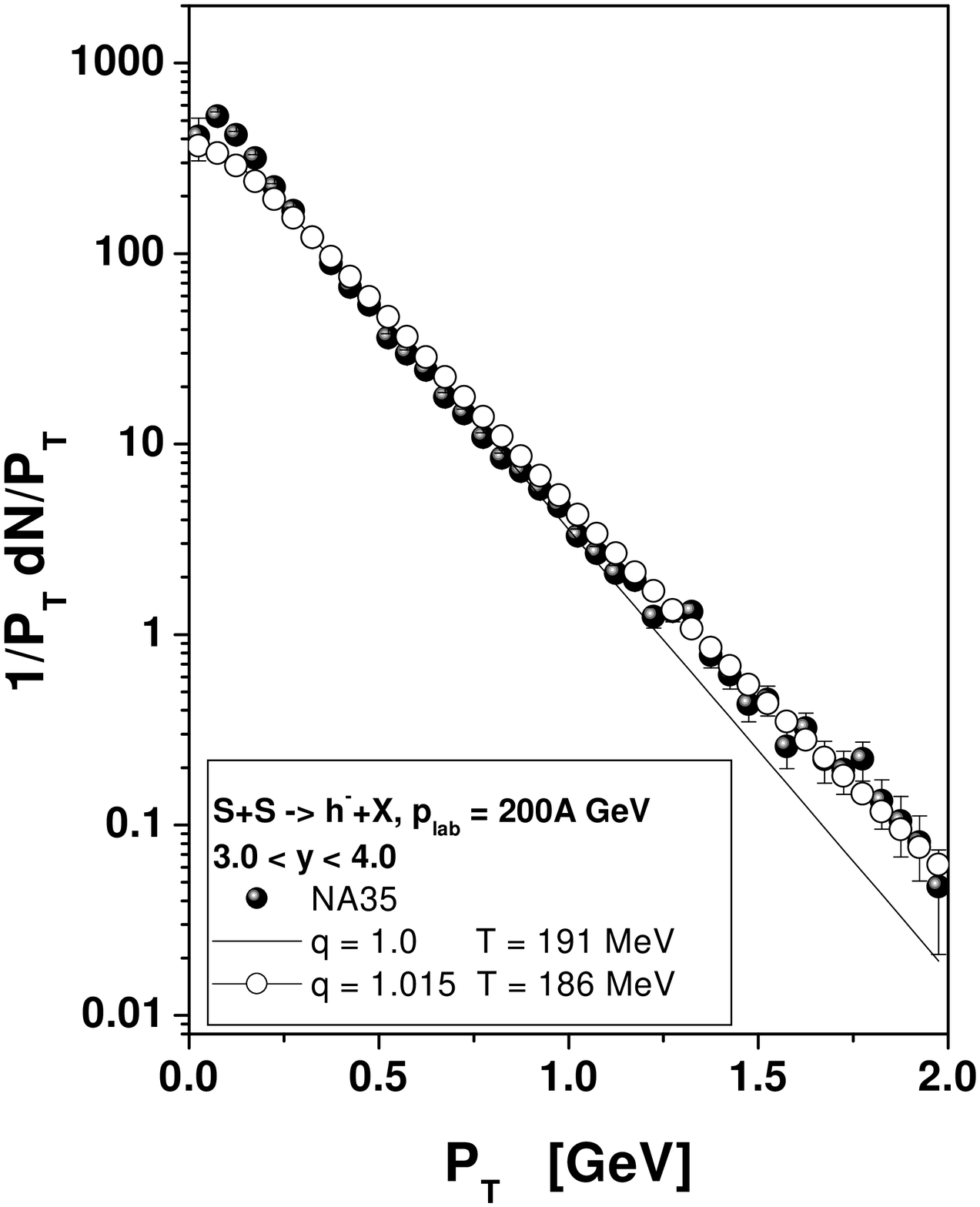, width=61mm}
     }
  \end{minipage}
  \vspace{-3mm}
  \caption{ \footnotesize {Transverse momentum spectra showing
nonextensive character. Left panel: fits to $p_T$ spectra from UA1
experiment on $\bar{p}p$ collisions \protect\cite{Datapt}. Here
$q=1.11$, $1.105$ and $1.095$ for energies decreasing from $900$ to
$200$ GeV, respectively (cf., \protect\cite{Kq}). Right panel: fit
to $p_T$ spectra from $S+S$ collisions \protect\cite{SS} (cf.,
\protect\cite{JPG}). Notice that in this case $q=1.015$ only, i.e.,
it is smaller less than for $\bar{p}p$ case, see eq.
(\ref{eq:fluct}).}}
 \label{fig:Fig3}
\end{figure}

Let us now proceed to examples of $p_T$ distributions, see Fig.
\ref{fig:Fig3}. Both $\bar{p}p$ data at high energies and nuclear
data at lower energies are shown. They all can be fitted only with
$q>1$ (for details see \cite{MaxEntq} and \cite{JPG}. The characteristic
feature is that now values of $q$ are much smaller than those
obtained fitting data in longitudinal phase space. We shall not
pursuit this problem here referring to discussion in \cite{MaxEntq}.
Instead we shall concentrate on the nuclear data (right panel) and
stress here that these data can, in our opinion, be connected with
fluctuations of temperature mentioned at the beginning. In fact the
$q=1.015$ obtained here fluctuation of $T$ of the order $\Delta T/T
\simeq 0.12$, which do not vanish with increasing multiplicity
\cite{WW}. In fact they are fluctuations existing in small parts of
the hadronic system with respect to the whole system rather than of the
event-by-event type, for which $\Delta T/T \sim 0.06/\sqrt{N}
\rightarrow 0$ for large $N$. Such fluctuations are potentially very
interesting because they provide a direct measure of the total heat
capacity of the system, $C$, because
\begin{equation}
\frac{\sigma^2(\beta)}{\langle \beta\rangle^2} \stackrel{def}{=}
\frac{1}{C} = q - 1. \label{eq:fluct}
\end{equation}
In fact, because $C$ grows with the volume $V$ of reaction we expect
that $q(hadronic) > q(nuclear)$ which seems to be indeed observed
(cf., for example, Fig. \ref{fig:Fig3}).
\vspace{-5mm}

\section{Summary and conclusions}

We have demonstrated that large amount of data on multiparticle
production can be quite adequately described by using tools from
information theory, especially when allowing for its nonextensive
realization based on Tsallis entropy. We have argued that
nonextensivity parameter $q$ entering here can, in addition to the
temperature parameter $T$ of the usual statistical approaches,
provide us valuable information on dynamical fluctuations present in
the hadronizing systems. Such information can be very useful when
searching for phase transition phenomena, which should be accompanied
by some special fluctuations of nonmonotonic character. It is
therefore worth to end with saying that some interesting fluctuations
of this sort seems to be already seen and investigated \cite{fluc}.
All that calls for some more systematic effort to describe existing
data in terms of $(T,q)$ for different configurations and energies in
order to find possible regularities in their system and energy
dependencies and possible correlations between them.

\section*{Acknowledgements} Partial support of the Polish State
Committee for Scientific Research (KBN) (grant 2P03B04123 (ZW) and
grants 621/E-78/SPUB/CERN/P-03/DZ4/99 and 3P03B05724 (GW)) is
acknowledged.

\end{document}